\documentstyle[prb,aps,twocolumn,amssymb]{revtex}
\begin{document}
\title{\bf Free particle scattering off two oscillating disks. }
\author{A. Antill\'on$^1$, Jorge V. Jos\'e$^{2,3}$ and  T. H. Seligman$^1$}
\address{$^{1}$ Instituto de F\'\i sica, Universidad Nacional Aut\'onoma
de M\'exico\\ Apartado Postal 48-3, 62251 Cuernavaca, Morelos, MEXICO\\
$^{2}$Physics Department and Center for the Interdisciplinary
Research on Complex Systems \\ Northeastern University, Boston, MA,
02115, USA$^\dagger$\\  $^{3}$Instituto de F\'{\i}sica,
Universidad Nacional Aut\' onoma de M\' exico,\\
Apartado. Postal 20-364, 01000 M\' exico D. F., M\' exico}

\maketitle
\begin{abstract}
We investigate the two-dimensional classical dynamics of the
scattering  of point particles by two periodically oscillating disks.
The dynamics exhibits regular and chaotic scattering properties,
as a function of the initial conditions and parameter values of the
system. The energy is not conserved since the particles can gain and
loose energy from the collisions with the disks. We find that for
incident particles whose velocity is on the order of the oscillating disk 
velocity, the energy  of the exiting particles displays non-monotonic 
gaps of  allowed energies, and the distribution of exiting particle 
velocities shows significant fluctuations
in the low energy regime. We also considered the case when the initial
velocity distribution is Gaussian, and found that for high energies
the exit velocity distribution is Gaussian with the same mean and
variance. When the initial particle velocities are in the irregular
regime the exit velocity distribution is Gaussian but with a smaller
mean and variance. The latter result can be understood as an example of
stochastic cooling. In the intermediate regime the exit velocity
distribution differs significantly from Gaussian. A comparison of the results
presented in this paper to previous chaotic static scattering problems is
also discussed.
\end{abstract}


\section{\bf Introduction }

The study of nonlinear systems capable of exhibiting chaotic behavior
has been an intensive area of research in the last fifteen years. 
This research was initiated
about a  century ago by the work of Henri Poincar\'e,  who studied
the motion of three gravitationally interacting bodies.
Most of the work done in this subject has focused on bounded
systems. On the other hand, many experimental techniques involve 
scattering processes.
In contrast to bounded systems, where
the particle's trajectories remain forever inside the range of interaction,
in a scattering process an incoming particle feels the interaction 
potential only for a finite amount of time and eventually exits the 
interaction region 
\cite{chat-intro,5a,6a,1a,2a,3a,BGO,7a,8a,9a,10a}.
In the general description of a scattering process, we have an input trajectory
into a region  of nontrivial dynamics called the scattering region,
and an output trajectory away from this region. We can think of the scattering
process as a map that transforms an incoming trajectory into an outgoing
one.  Only relatively recently it has been realized that a scattering
processes from a general scattering potential, often without a simple
geometric symmetry, can have rather complicated dependencies between the 
incoming and outgoing trajectories. This means that, by very slightly changing 
the initial conditions that define the incoming trajectory,  the outgoing 
one will  have rather large fluctuations. The idea that {\it chaotic 
scattering}  can play an important
role in various problems in physics became widely accepted
after the seminal work of Petit and H\'enon \cite{1a,chat-intro}.

Most previous chaotic scattering studies have assumed a stationary
scattering region, i.e. fixed in time (for an exception see \cite{11a}).
In this paper we present results from a dynamical study 
of the scattering of particles from a time-dependent oscillatory
interaction potential, which consists of two circular disks
that oscillate periodically in time. The static two-disk problem 
was recently shown to be analytically integrable \cite{10a},
(hereafter we call this work I). In this paper we build our nonequilibrium 
dynamical study based upon the results obtained in I.

Our model can conceivably be produced in very low temperature experiments 
where a couple of circular quantum dots are generated by a gate voltage
that can vary their radius periodically in time. 
Ballistic transport experiments in mesoscopic systems  
have raised the  possibility of directly studying  chaotic billiards, 
where the addition of external fields can yield results 
that are expected to account for certain aspects of unusual related 
experimental 
results \cite{ball,antidot,antidot-2,1,2,3,4,5}.
Some of the transport results seen in experiments are surmised
to have classical related explanations \cite{BH9,RA,Baskin,Fliesser,GKS,FGK}.
The geometry of microjunctions
\cite{ball} and antidot-lattices \cite{antidot,antidot-2} can be
described by models that consist of circular scattering  disks.
For the above reasons, we will focus in this paper on the scattering 
of a particle from two 
oscillating classical hard-disk billiards. Here we concentrate
on the classical dynamics of this model and leave the very 
interesting quantum case for a future study.

The outline of the paper is the following:
In  Section II we introduce the model considered in this paper,
together with its main physical properties. In section III we  derive a  
scattering map associated with our problem. In section IV we present 
and discuss the bulk of our results. Finally, in section V we provide a 
short summary  of the results and the perspectives for the future.

\section{\bf Definition of the problem.}

We consider the motion of a unit mass  particle restricted to
move on the plane. The particle elastically  collides  with two hard 
disks that oscillate periodically in time.
The initial velocity of the particle changes as a function of time
due to the energy exchange after each collision with the disks.
As we discuss below, depending on the initial conditions the particle
will spend a certain amount of {\it dwell time} in the interaction 
region, after which time it will exit upwards or downwards. It is the 
complexity  of this motion that we will carefully describe below.

Here we will follow the approach presented in I \cite{10a}, including its 
notation. The reader  should check this reference for further details on 
the formulation of the static problem. In Fig. (\ref{fig1}) we show the two
disks on the plane. The radius of both disks are normalized to one. Their
centers are separated by a time-dependent distance $R(t)>2$. One convenient
way to study this problem, as pointed out in I, is by
replacing the system by one disk and one rigid wall placed at
the symmetry axis of the two-disk problem. This is the representation of
the model  we study in this paper.

\subsection{ \bf Two-disk oscillating model.}

The model we consider here is in some sense the scattering  two-dimensional 
extension  of the well studied {\it bound Fermi acceleration} model 
\cite{12a,13a}. This model is  defined by a free particle inside a rigid 
one-dimensional box, with one wall fixed and the other one periodically 
oscillating in time.  The Fermi model was one of the first two degrees of
freedom problems studied, which exhibited a transition 
from regular to chaotic behavior
 as a function of the oscillating wall motion. For a linear saw--tooth 
time--dependent wall oscillation, 
the particle dynamics is regular. Having a linear time--dependence implies
a constant oscillating wall velocity.  When the oscillation
is nonlinear in time, there is acceleration in the wall motion and one can
then have non-trivial dynamics, with a transition between regular 
to fully chaotic behavior. In this paper, without loss of generality, 
we consider the simplest nonlinear piece--wise quadratic time-dependent 
disk oscillation shown in Fig. (\ref{fig2}).
In this case we represent the  motion of the disk center by 
\begin{equation}
X(t)=\tilde At^2+\tilde Bt+\tilde C.
\label{one}
\end{equation}
Here the constants $\tilde A$, $\tilde B$ and $\tilde C$ are fixed for a
half period, and they have different values for  different  half
periods.  The motion of the disk center  (the other disk
is the mirror image of this one) is given by Eq. (\ref{one})
modulus the oscillation disk period $T$.

In our analysis, for calculational convenience, we chose to 
treat the problem 
in the following way.  We  label by the integer $m$ each continuous 
piece of the disk oscillation. The time $t$ and $m$ are then related 
by the expression
\begin{equation}
m=\left\lbrack\!\left\lbrack{2(t+\Phi_0/\omega)\over
T}\right\rbrack\!\right\rbrack,
\label{two}
\end{equation}
where $\left\lbrack\!\left\lbrack\ \
\right\rbrack\!\right\rbrack$ denotes the nearest lower integer. The
$m$ parameter will have a fixed value for time $t$ in the time interval
$({m\over 2})T\le t \le ({m+1 \over 2})T$.
Here $\omega=2\pi/T$ is the oscillation frequency, and $\Phi_0$ is the initial
oscillation phase. The specific  expressions that define the parameters
$\tilde A$, $\tilde B$ and $\tilde C$ are given in Appendix A.

We have now defined the time dependence of the oscillating disk. Next we
use the relevant results  given in I, noting that 
the incidence--reflexion symmetry in our case is changed by the oscillation
of the disk.

\subsubsection{\bf Collision Time}

We start by calculating the time elapsed between two successive collisions
of the particle with the disk.  We need this time to calculate the  
new velocity vector,  by means of a velocity transformation to the 
system where the disk is at rest.
We deduce from Fig. (\ref{fig1}) that the position of the colliding 
particle is  given by
\begin{equation}
\vec \rho(t_{n+1})=\vec\rho_{n+1}=\vec \rho_{nv}+\vec v_{nv}(t-t_n),
\label{three}
\end{equation}
where $t_n$ is the previous collision time, and the subindex $v$ 
denotes a {\it specular variable}. To clarify the meaning of {\it specular},
consider for example, the one associated with $\vec \rho_n(t)=(x_n,y_n)$ 
which gives
\begin{equation}
\vec \rho_{nv}=(x_{nv},y_{nv})=(-x_n,y_n)=(-X_n+\cos\theta_n,\sin\theta_n),
\label{four}
\end{equation}
and the velocity
\begin{eqnarray}
\vec v_{nv}&=&(v_{nvx},v_{nvy})=(-v_{nx},v_{ny})\nonumber\\
& =&(v_n\cos(\theta_n-\phi_n),
v_n\sin(\theta_n-\phi_n)).
\label{five}
\end{eqnarray}
The new collision point $\vec\rho_{n+1}=(x_{n+1},y_{n+1})$, 
at the new collision time $t_{n+1}$,  must lie on the circumference 
given by the equation 
\begin{equation}
(x_{n+1}-X(t_{n+1}))^2+y_{n+1}^2=1.
\label{six}
\end{equation}
Evaluating Eq. (\ref{one}) at $t=t_{n+1}$, and substituting it in 
Eq. (\ref{six}), we get the quartic  equation for $t_{n+1}$
\begin{equation}
a_4t^4_{n+1}+a_3t^3_{n+1}+a_2t^2_{n+1}+a_1t_{n+1}+a_0=0,
\label{seven}
\end{equation}
where the expression for the parameters $a_0$, $a_1$, $a_2$, $a_3$ and $a_4$,
are explicitly given in Eq. (A.2).

We can get  $t_{n+1}$ as a  function of $t_n$ using Eq. (\ref{seven}) and
Eq. (A.2). Once we know $t_{n+1}$, the collision point 
on the disk can be determined  from Eq. (\ref{three}) as
\begin{equation}
x_{n+1}=-x_n-v_{nx}(t_{n+1}-t_n),
y_{n+1}=y_n+v_{ny}(t_{n+1}-t_n),
\label{eight}
\end{equation}
and then, using Eq. (\ref{six}), the disk will be located at
\begin{equation}
X_{n+1}=x_{n+1}+\sqrt{1-y_{n+1}^2}
\label{nine}
\end{equation}

\subsubsection{\bf {Disk velocity Map.}}

To calculate the  velocity of the disk, $\vec V_{n+1}$, at the new
collision time, we take the time-derivative of Eq. (\ref{one}) 
that gives
\begin{equation}
\vec V(t)=(\dot X(t),0)=(2\tilde At+\tilde B,0),
\label{ten}
\end{equation}
and consequently
\begin{equation}
\vec V_{n+1}=\vec V(t_{n+1})=(2\tilde At_{n+1}+\tilde B,0),
\equiv(V_{n+1},0)
\label{eleven}
\end{equation}
which is fully determined since $t_{n+1}$ is  known from Eq. (\ref{seven}).
To determine  the velocity of the particle, $\vec v_{n+1}$, we introduce  
the relative  particle velocity (see Fig. (\ref{fig3})) 
with respect to the disk as
\begin{equation}
\vec u=\vec v-\vec V.
\label{twelve}
\end{equation}
 Then 
\begin{equation}
\vec u_{nv} \wedge  
\hat r_{n+1}=P\hat k,
\label{therteen}
\end{equation}
where
\begin{equation}
P=((X_{n+1}-x_{n+1})u_{ny}-y_{n+1}u_{nx}),
\label{fourteen}
\end{equation}
and 
\begin{equation}
u_{nx}=v_{nx}+V_{n+1}\qquad ; \qquad u_{ny}=v_{ny},
\label{fifteen}
\end{equation}
with the normal unit vector $\hat k \parallel \hat z$,  as seen in the
figure. One can also show that
\begin{equation} 
\vec u_{n+1}\wedge\hat r_{n+1}=(u_{n+1x}y_{n+1}-
u_{n+1y}(x_{n+1}-X_{n+1}))\hat k.
\label{sixteen}
\end{equation}
From these two equations we get
\begin{equation}
u_{n+1y}=\bar P-Qu_{n+1x},
\label{seventeen}
\end{equation}
with 
\begin{equation}
\bar P={P\over X_{n+1}-x_{n+1}}\qquad \hbox{and} \qquad Q={y_{n+1}\over
X_{n+1}-x_{n+1}}.
\label{eighteen}
\end{equation}
We also have the conservation of the velocity magnitude, which in the coordinate
frame where the disk is at rest is given by 
\begin{equation}
u_{n+1x}^2+u_{n+1y}^2=u_n^2.
\label{nineteen}
\end{equation}
Using Eq. (\ref{fifteen}) in the last equation we get
\begin{equation}
u_{n+1x}={\bar PQ-\sqrt{\bar P^2Q^2-(1+Q^2)(\bar P^2-u_n^2)}\over
1+Q^2},
\label{twenty}
\end{equation}
with an appropriately chosen $``-"$ sign in the square root. 
We now  go back  to Eq. (\ref{fifteen})
to find $u_{n+1y}$. This allow us to obtain the velocity of the particle
after the collision through the expressions
\begin{equation}
v_{n+1x}=u_{n+1x}+V_{n+1}\qquad \hbox{and} \qquad v_{n+1y}=u_{n+1y}.
\label{twentyone}
\end{equation}

\section{\bf The scattering map.}

Following the notation of I, we can get the scattering map associated with 
this dynamical system.  Since the derivation of our map is completely
analogous to the one given there, we can  directly write down the
final expressions stating a few differences proper to our problem. 
The map derived in I is
\begin{equation}
\phi_{n+1}=\sin^{-1}\left[{v_n\over v_{n+1}}(\sin\phi_n+\bar R
\sin(\theta_n-\phi_n))\right],
\label{twentytwo}
\end{equation}
\begin{equation}
\theta_{n+1}=\sin^{-1}(\sin\theta_n+\lambda\sin(\theta_n-\phi_n)).
\label{twentythree}
\end{equation}
In our case $\bar R=X_n+X_{n+1}$ and
\begin{eqnarray}
\lambda&=&\bar R\cos(\theta_n-\phi_n)-\cos\phi_n\nonumber\\
 &-& \sqrt{[\cos\phi_n-\bar R\cos(\theta_n-\phi_n)]^2-\bar R^2 
+2\bar R\cos\theta_n}.
\label{twentyfour}
\end{eqnarray}
The initial conditions  $(\theta_0,\phi_0)$  in this map are obtained
from the parameters set at time $t=0$. If we take the initial particle position
as $(x_{\bar 0},y_{\bar 0})$, the angle $\alpha$ between the initial 
velocity $\vec v_{\bar 0}$ and the horizontal gives 
\begin{equation}
\phi_0=\sin^{-1}\left[{v_{\bar0}\over
v_0}((X_0-x_{\bar0})\sin\alpha +y_{\bar0}\cos\alpha)\right],
\label{twentyfive}
\end{equation}
\begin{equation}
\theta_0=\sin^{-1}(y_{\bar0}+\lambda_{\bar0}\sin\alpha).
\label{twentysix}
\end{equation}
In Eq. (\ref{twentyfive}), $\vec v_0$ is given by 
$\vec v_0=\vec u_0+\vec V_0$, and  $\vec u_0$ by
\begin{equation}
u_{0x}={\bar RS-\sqrt{\bar R^2S^2-(1+S^2)(\bar R^2-u_{\bar 0}^2)}\over
1+S^2},
\label{twentyseven}
\end{equation}
\begin{equation}
u_{0y}=\bar R-Su_{0x},
\label{twentyeight}
\end{equation}
where
\begin{equation}
\bar R={R\over X_0-x_0}\qquad \hbox{and} \qquad S={y_0\over
X_0-x_0},
\label{twentynine}
\end{equation}
\begin{equation}
R=(X_0-x_0)u_{\bar 0y}+y_0u_{\bar 0x}\qquad \hbox{with}\qquad u_{\bar
0x}=v_{\bar 0x}-V_0\qquad \hbox{and} \qquad u_{\bar 0y}=v_{\bar 0y}.
\label{thirty}
\end{equation}
The results derived in this section used a polar coordinates representation
that has several advantages for the geometric analysis described here. 
When iterating the map numerically, however, the polar coordinate
representation is somewhat cumbersome and for that reason we found it
more convenient to carry out the iterations
in Cartesian coordinates. This is what we did to obtain the results described 
in the next Section.

\section{\bf Results }

In this Section we discuss the bulk of our numerical results. We provide
typical results for a regime of interesting physical parameters.
To check our analysis, we looked at the Fermi acceleration limit 
of our problem, which corresponds to having the two disks quite close
to each other and with the particle initial conditions along the disks axis,
so that the particle does not notice the disks curvature.
We reproduced the Fermi accelerator model results  by choosing 
the parameters for the equilibrium position of the disk center, $X_e$, 
the amplitude of oscillation, $A$, the time oscillation period 
 $T$, and the free space distance between the wall and the disk, 
close  to the values given in Refs. \cite{12a,13a}. 
The  phase space plots obtained correspond well to the
known Fermi accelerator results, i.e. chaotic behavior for low 
velocities, and several sets of resonant islands for higher velocities. 

After this test we proceed to chose the separation between the wall and disk 
large enough so that the particle dynamics sensed the curvature of the disks.
Of course, if the separation distance is too large the particle will hardly 
collide with the disk and the dynamics becomes trivial. The interesting
parameter ranges are the ones which allow  a large number of particle 
collisions with the oscillating disk and the wall.
A typical phase space plot is shown in 
Fig. (\ref{fig4}), for several particle initial conditions.
The parameters considered satisfy the necessary condition to have a large 
number of particle collisions with the oscillating disk. 
In the units where the  radius of the disk is one, we took the parameters: 
$(X_e,0)=(1.000097,0)$, $A=1.6\times 10^{-6}$, $T=7.6\times 10^{-5}$, 
$\Phi_0=0$, and the  acceleration parameter $\tilde A=4716.085$. 
We considered a set of 2000  particle initial conditions, 
that we may  as well call a beam  of 2000 particles, each one sent from 
the  origin into  the  scattering region with an angle 
$\alpha=6\times  10^{-2}$ radians with respect to the x--axis. 
We  varied  the velocities of the particles between 0.1-5.0, chosen from
a uniform random distribution. 

In Fig. (\ref{fig5}) we show the delay, or {\it dwell time}, $\tau_d$, 
as a function of the initial energy (velocity) of the incident particles. 
For low energies we observe a very irregular behavior in $\tau_d$.  
In fact, this behavior is rather close to a fractal, as can be seen 
from zooming  in a given interval of energies, as shown in the left 
inset. We also note that for initial velocities larger that $5.5$, 
there is a mixture of regular and irregular zones. When we amplify 
one of the irregular regions, we can again see the fractal character 
of the results (see inset on the right). For larger energies than 
the ones shown here, we found that $\tau_d$ tends to a constant value. 
This is what we would expect for large  energies since the particle 
essentially sees the disk as stationary.

It is interesting to use the data of Fig. (\ref{fig5}) to construct the
histogram of  dwell times shown in Fig.  (\ref{fig6}). The main 
histogram has two representative  contributions. One comes 
from the irregular zone and the other one from the semiregular component, 
as it is shown in the two insets in the  figure. The upper inset
corresponds to the irregular region and the lower one to the 
semiregular zone. The main histogram shows one peak close to a 
$\tau_d$  of about $100$  and the other one close to $200$,  
which correspond to the peaks seen in the insets. The bin size 
used in all the histograms shown are around $3\%$ of the full range.
In Fig.  (\ref{fig7}) we show the number of collisions with the disk
 versus the incoming velocity. The general behavior is very similar to that
in Fig.  (\ref{fig5}), however, the explicit relationship  is complicated.
This can also be seen from   Fig. (\ref{fig8}), which we shall  discuss in the
next paragraph. In  Fig.  (\ref{fig7}) we note, in particular, the 
irregular behavior in the same regions of incident velocities. 
As we increase the initial energy, the number of collisions reach a 
plateau, which  is because the disk appears to be at rest. 
In the inset we used the same number of particles 
as used in the main figure, but the range of velocities is smaller,
so as to allow us  to see in more detail the results.

In Fig.  (\ref{fig8})  we display the dwell time vs the number 
of collisions, $N$, to see if there is a simple relation between them.
The input velocity 
appears here only as a hidden variable. For example, when 
the range of velocities 
in the inset is in the irregular region, there is a wide  spread 
in the location of the resulting points. If we also 
allow  initial velocities from 
the semiregular region, as it happens  in the main figure, 
then the data points are localized around a specific zone that is darker 
in the figure. This is consistent with an essential
independence of these variables when the initial 
velocities are outside of the irregular region.  
The number of collision data points in the 
main figure and in the inset are equal.

Next we  discuss the relevant scattering variables of the problem. 
In Fig.  (\ref{fig9})  
we show the irregular behavior of the exit angle as a function of 
the normalized initial particle  velocity. The low energy 
particles, with velocities less than $5.3$, get an irregular
exit angle in a wider range of values. When we plot the distribution of 
these exit angles, we find a wide pattern centered
around an angle of about 0.45 radians. This can be seen in the inset of this 
Fig. (\ref{fig9}). Particles with incident velocities larger than $5.3$,
that show semiregular behavior,  also contribute strongly to this peak. 
The  corresponding histograms for these regions are displayed in 
Fig. (\ref{fig10}), with the left histogram associated with the 
irregular region, and the right one with  the semiregular region. 
Both histograms show a peak for an exit angle  around $0.45$ radians. 
For low energy incident particles we get  a wider range  of output angles.

In Fig. (\ref{fig11})   we show the exit velocity as function of 
 input velocity. The region with input velocity less than $5.3$ 
is quite irregular, and it is consistent with the previous figures. 
When we increase the input velocity, the exit velocity grows and 
the fluctuations are about a line with slope of 
almost $45^\circ$.  In this figure it is noticeable that 
there are big jumps for input velocities 
between $16.7$ to $19$. The left inset is an 
amplification for low input velocities. The right inset shows the 
corresponding histogram for the exit velocities. Here we notice that 
there are isolated peaks or gaps in this distribution.

The histograms shown in Fig. (\ref{fig12}) are directly related to 
the  data shown in Fig. (\ref{fig11}). The left histogram corresponds to 
the data given in the inset of  Fig. (\ref{fig11}), while the right one  
corresponds to the main plot.  In both analyzed histograms the fluctuations 
are about the line with slope $\pi/4$, with data points on this line  
labeled by the variable $v$. The variables $V_{max}$ and $v_{out}$ are the 
maximum disk velocity and the real output velocity of the particles,
respectively. One 
important feature of these histograms is that they appear to be 
directly related
to the isolated peaks  mentioned before. Here the effect is more prominent 
for the irregular region of  input velocities. The effect remains, even if we 
make the  velocity equal, but with different initial phases. 
This is equivalent to carry out a phase average  
in the interval $(0,2\pi)$. The gaps in the output velocities 
are also gaps in output energy. These results  indicate that there are 
 output energy regions that the particle can not explore, leading
to {\it forbidden} energy regions.
In the inset at the top of Fig.  (\ref{fig11}), 
we note that the exit velocities have a peak when the input 
velocities are close 
to $19$. For the range of velocities between $23-60$ the energy gaps
were not seen.  This is why the gaps are wider at the bottom in the 
inset of Fig.  (\ref{fig11}). If we increase the range of input velocities,
 as in the main figure, then there appear narrow gaps related to 
different velocity contributions. 

We have also carried out a  basic  fractal 
analysis of a 10,000 particle system. The idea was to extend the analysis
of Ref. \cite{2a} to  two-dimensions.  We determined 
the plane boundary  of initial conditions
$(x_{\bar0},\alpha_0)$, which separates the particles
into the ones that go upwards from the ones that go downwards.  We
plotted a figure with black squares representing the initial conditions 
of particles  which exit upwards, and empty  squares belonging to the 
ones that go downwards. We obtained $1.86$ as the fractal dimension.
We do not show these results since they are
typical of chaotic scattering problems. We carried out this quantitative
analysis to make sure that all the  qualitative generic properties of a 
chaotic scattering system applied. Although all  the results are quantitatively 
different, as one should expect,  we did not find a significant change in 
the  general qualitative behavior described above. 

Finally, we note that the model we are considering here does not conserve
energy, and we are also interested in understanding how energy is added
or subtracted from the disk to the colliding particles. One possibility
is to take the initial velocities distributed by a Gaussian function,
just as in the classical statistical mechanics Maxwell velocity distribution.
We chose then a beam of particles with this velocity distribution
with a given standard deviation $\sigma$, or inverse temperature.
Then we studied the evolution
of the distribution of exit velocities.  We did the analysis
at low, intermediate and high beam energies.  The results are
shown in Fig.  (\ref{fig13}). In all of these figures, 
the continuous line curve represents the Gaussian distribution fit
to the beam of incident particle velocities. The histogram  is the distribution 
of exit velocities after scattering.  We note that  
the Gaussian distribution is maintained {\it only} for high energies 
(right--bottom figure), but for low or medium energies the exit velocities
can be not be fitted to a simple Gaussian. At low energies, 
when the incident velocities are in the irregular region (left--top figure), 
however, most particles concentrate about a Gaussian-like 
distribution, with smaller mean and $\sigma$. These latter results indicate 
that the beam losses energy and that it has some kind of stochastic 
cooling.

\section{ Conclusions. }

In the present paper we considered the complex 
dynamics of a particle that scatters from two periodically oscillating disks
with a variety of initial conditions.
We found that the dynamics has regular and irregular behavior that we analyzed
in some detail. 
This model is in a sense a {\it dynamical extension} of 
the well known Lorentz gas \cite{lg0}. Although the model studied here
is perhaps  the first dynamic chaotic scattering analysis,
 several of the results we described 
are similar to those found in chaotic static scatters.  There are, however, 
some important unexpected differences in the results obtained. 
Among the  most interesting 
and surprising results presented in this paper are the  energy gaps 
found in the exit energy. This result indicates that there is an
important  energy absorbing mechanisms, 
directly related to the nature of the classical dynamics of the problem. 
We found that the markedly irregular dynamics appears when the particles 
have velocities on  the order of the disk velocities. It is  within 
this energy range  that the energy gaps appear.  For larger particle 
energy the dynamics simplifies, for the  oscillating disks  appear as if 
they were at rest.

Another important difference from the dynamics of chaotic
static scatterers, has to do 
with the energy gained or lost by the beam of particles. As a test,
we took a Maxwell--like distribution of initial velocities and 
found that only in the large energy regime, where the disk is essentially 
seen at rest by the particles, the Gaussian 
distribution of exit velocities is preserved. Otherwise,
there are important changes in the exit velocity distributions
for low velocities.

In this paper we considered that the disks do not absorb energy
from the colliding particles. Including energy gained or lost by the
disks is necessary to mimic the effect of temperature in the model.
We intend to include this effect in the model elsewhere.  The very
interesting questions raised by the quantum mechanical treatment
of the model introduced here are left for the future.

\section*{Acknowledgments}

The work by AA was partially supported by DGAPA-UNAM grant IN-105595,
and the one by JVJ was supported in part by NSF grant DMR-95-21845.

\newpage
\appendix

\section*A
In this appendix we write down the explicit expressions for the
parameters defined in the main body of the text.
In order to obtain the values of the parameters $\tilde A$, $\tilde B$
and $\tilde C$ of section II, we use Fig. (\ref{fig2}), and ask  
that the parabolic  curve crosses trough the points 
$({m\over 2}T,X_e+A)$ and $({m+1\over 2}T,X_e-A)$,
where $X_e$ is the equilibrium position of the center of the disk and
$A$ is the oscillation amplitude. When this is done, we obtain the 
following expressions:

\begin{eqnarray}
\tilde A&=&\hbox{free parameter },\nonumber \\
\tilde B&=&(-1)^m{2A\over\pi}+\left({\tilde A\over \omega}\right)
\left(2\Phi_0-\pi(1+2m)\right), \nonumber \\
\tilde C&=&X_e+(-1)^{m+1}A(1+2m)+
(-1)^m{2A\Phi_0\over\pi}\nonumber\\
&+&\left({\tilde A\over
\omega^2}\right)\left(\Phi_0^2+m\pi^2(1+m)-\pi\Phi_0(1+2m)\right).\nonumber
\end{eqnarray}

$$\eqno(A.1)$$
$\tilde A$ is the curvature in the saw tooth, which is associated to the
disk acceleration.

The derivation of the parameters $a_0$, $a_1$, $a_2$, $a_3$ and $a_4$
appearing in Eq. (\ref{seven})
is a straightforward and here we just cite the results. We 
evaluate Eq. ({\ref{one}) at $t=t_{n+1}$ and then substitute the result into 
Eq. (\ref{six}). Then  we obtain Eq. (\ref{seven}) with the 
following coefficients.

\begin{eqnarray}
a_0&=&x_n^2+v_{n}^2t_n^2+\tilde C^2-2x_nv_{nx}t_n+2x_n\tilde
C\nonumber\\
&-&2v_{nx}\tilde Ct_n+y_n^2-2y_nv_{ny}t_n-1, \nonumber \\
a_1&=&-2v_{nx}^2t_n+2\tilde B\tilde C+2x_nv_{nx}+2x_n\tilde
B\nonumber\\
&-&2v_{nx}\tilde Bt_n+2v_{nx}\tilde C-2v_{ny}^2t_n+2y_nv_{ny},\nonumber \\
a_2&=&v_n^2+\tilde B^2+2\tilde A\tilde C+2x_n\tilde A-2v_{nx}\tilde
At_n+2v_{nx}\tilde B ,\nonumber \\
a_3&=&2\tilde A\tilde B+2v_{nx}\tilde A, \nonumber \\
a_4&=&\tilde A^2 .\nonumber
\end{eqnarray}
$$\eqno(A.2)$$

\newpage

\vskip 1cm
\begin{figure}[hp]\begin{center}
\caption{
This figure defines the two-disk periodically oscillating model 
studied in this paper. The model is replaced and studied by the one oscillating
disk and a fixed wall. The figure shows a dotted line disk with 
center at $X_n=X(t_n)$, and a continuous line circle at the 
right of the wall with  center at $X_{n+1}=X(t_{n+1})$. 
The disk on the left of the wall represents the image of the dotted line disk.
Variables with a subindex $n$ are evaluated at time $t_n$.
\label{fig1}}
\end{center}\end{figure}

\begin{figure}[hp]\begin{center}
\caption{
Here $X(t)$ denotes the  oscillating wall model 
about  the equilibrium position $X_e$ in the range  $[X_e-A,X_e+A]$, 
with amplitude $A$.   The parameter $m$ labels the oscillation 
segment with fixed  value in the time interval
$({m\over 2}T,{m+1\over 2}T)$. The relation between time $t$ and $m$ 
is given in Eq. (2). The figure is drawn for $\Phi_0=0$.
\label{fig2}}
\end{center}\end{figure}
\begin{figure}[hp]\begin{center}
\caption{
This figure is used in the derivation of the velocity map.  The relative 
velocity before  and after the collision has the same angle with respect 
to the normal  to the disk, but the velocity itself has two different angles.
See text for the definition of the variables in this figure.
\label{fig3}}
\end{center}\end{figure}

\begin{figure}[hp]\begin{center}
\caption{
Phase space results for a limit close to the Fermi acceleration 
one-dimensional model. Here $v_x$ is the x component of the
particle velocity, normalized  by the disk velocity $V_{max}$.
 The particle will eventually sense the disk 
two-dimensional curvature, and the resonances structure, as compared
to the Fermi model,  will change.  The plot has 74238 points obtained 
from $17$ particles with different initial velocities.
\label{fig4}}
\end{center}\end{figure}

\begin{figure}[hp]\begin{center}
\caption{
Dwell time $\tau_d$ as a function of the normalized incident velocity
$v_{in}$. In inset $a$ we show amplified results about $v_{in}=2.375$.
We see that the low energy particles have an irregular behavior with a 
fractal-like character. Velocities higher than $5.5$ have a mixture 
of regular and irregular behavior. In inset $b$ we show us a further 
amplification around velocities close to $7.581$. Each picture is drawn
from results of 2000 initial conditions.
\label{fig5}}
\end{center}\end{figure}

\begin{figure}[hp]\begin{center}
\caption{
Dwell time histogram $N(\tau_d/T)$ for data like the one shown in Fig. (5), 
except that here we used  6000  particles instead of the 2000 used in
Fig.  (5). The two  noticeable peaks are due to the contributions from  the
irregular and semiregular zones,  which correspond to the data shown in
insets $a$ and $b$, respectively. The number of  particles used to 
obtain the histograms was 2000.
The bin size in the three histograms is  around 3\% of the 
full range in each plot.
\label{fig6}}
\end{center}\end{figure}

\begin{figure}[hp]\begin{center}
\caption{
Here we show the dependence between $N$  and $v_{in}/V_{max}$.
The main plot and the inset show the same general behavior as that 
of Fig. (5), with the  irregular behavior in the same range of 
input velocities. For large velocities, 
$N$ is basically  constant. This  means that the disk motion 
has little effect in the particle velocities. Each picture was done with 
2000 particles.
\label{fig7}}
\end{center}\end{figure}

\begin{figure}[hp]\begin{center}
\caption{
Here we show a plot of $\tau_d$  vs  $N$.
We note that there is no clear relationship between these two variables.
The  general behavior of both plots are  similar to the
ones shown in Figs. (5) and (7).  Note  that for low $v_{in}$ 
(irregular region), their dependence is irregular. 
The number  of points taken was 2000. We see that when the input 
velocity is outside the irregular region, all the particles 
concentrate in a black zone in the figure. This means the almost
independence of $\tau_d$ and $N$ with respect to the input velocity.
\label{fig8}}
\end{center}\end{figure}

\begin{figure}[hp]\begin{center}
\caption{
Exit angle $\alpha_{out}$  as a function of $v_{in}/V_{max}$.
The low velocity region again  displays an irregular behavior for the
exit angle. The inset in the figure for the preferred exit angle 
has a peak around $0.45$ radians.
\label{fig9}}
\end{center}\end{figure}

\begin{figure}[hp]\begin{center}
\caption{
Histograms of the data $N(\alpha_{out})$  shown in the inset of Fig. (9) 
that separate irregular from semiregular regions. Both histograms
have a peak close to $\alpha_{out}=0.45$ radians. The irregular region
histogram is shown in $a$. It covers a wider range of output  
velocities than the semiregular region shown in histogram $b$. 
The bin sizes taken for the histograms are $3\%$ of the full range 
in $\alpha_{out}$.
\label{fig10}}
\end{center}\end{figure}

\begin{figure}[hp]\begin{center}
\caption{
In the main figure we show the exit velocity vs input velocity for
 6000 particles for a wide range of initial velocities.
Inset $a$ has the same coordinates, but for 2000 input velocities.
In inset $a$ the irregular region is more detailed, and between 0-15
the coarse averaged slope is small, while outside this range the
averaged slope is close to $\pi/4$. We use the latter result in the
histogram of Fig. (12).
The exit velocity histogram is shown in inset $b$. 
Notice the isolated peaks in the distribution.
\label{fig11}}
\end{center}\end{figure}

\begin{figure}[hp]\begin{center}
\caption{
Histogram of the $N(v_{out}-v)/V_{max}$ data.  $a$ 
corresponds to the data given in inset  of Fig. (11)(a), and $b$ to
the data shown in the main plot. The fluctuations are analyzed around
the line with slope $\pi/4$, and  are labeled by the variable $v$.
Here $v_{out}$  is the
real output velocity of the particles. We surmise that some of the 
isolated peaks present may be connected with the exit energy gap regions.
\label{fig12}}
\end{center}\end{figure}

\begin{figure}[hp]\begin{center}
\caption{
Each of the four figures was produced from a beam of particles with a
Gaussian initial velocity distribution, denoted by a continuous line
in the figures. The histograms of $N(v_{in})$ or $N(v_{out}$ 
were produced from the exit velocities after scattering the oscillating disks.
Each figure has a different mean and 
standard deviation. In  figure $d$ we see that the Gaussian  distribution is
preserved but only for high energies. In the irregular exit velocity
region ($a$), the exit distribution is concentrated about a Gaussian-like
distribution with smaller mean and $\sigma$. This result can be
interpreted as some kind of stochastic cooling mean and $\sigma$. This
result can be interpreted as some kind of stochastic cooling
Figures ($b$) and ($c$) correspond to intermediate energy regimes where
the exit distribution is not Gaussian.
\label{fig13}}
\end{center}\end{figure}

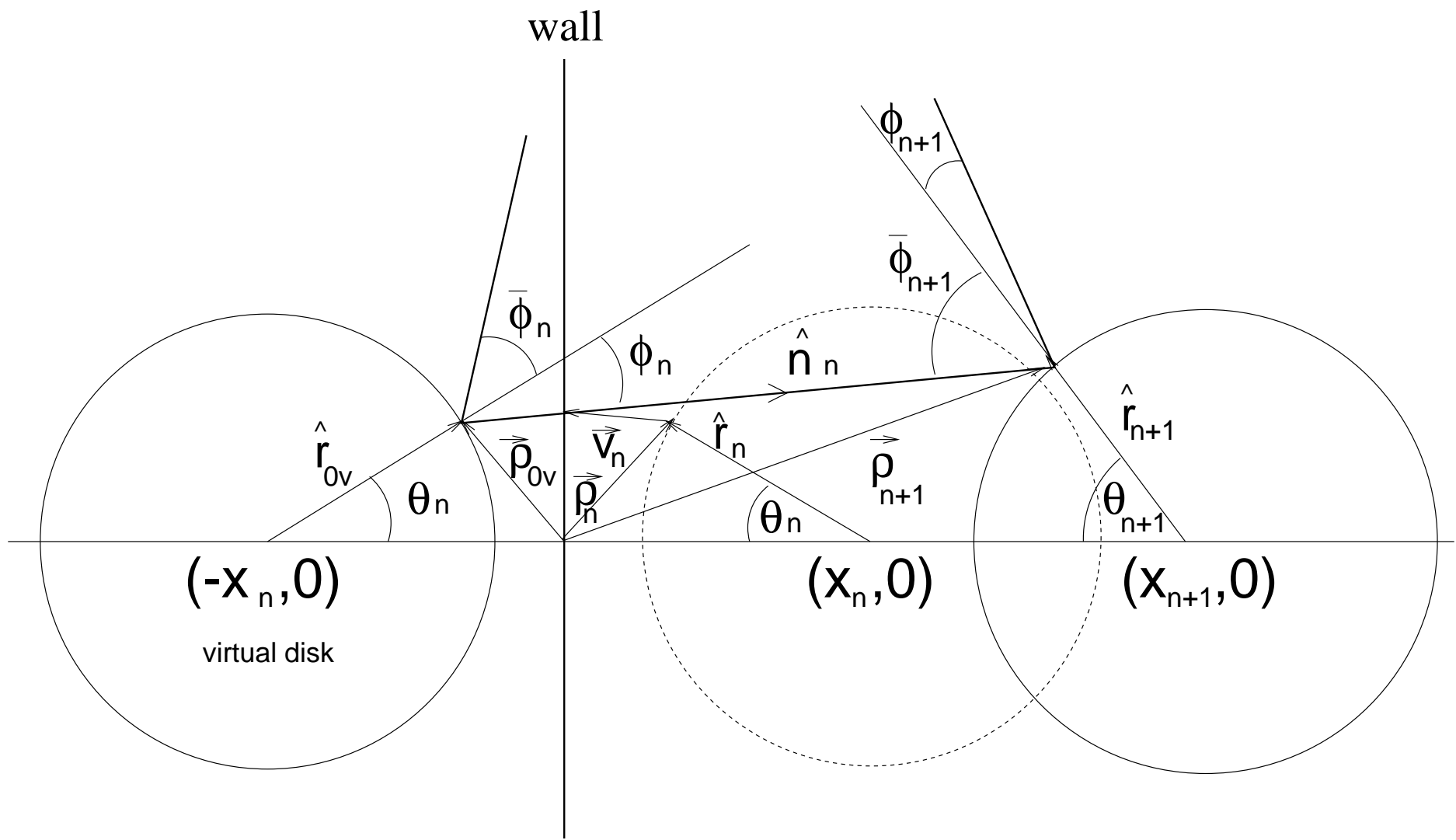
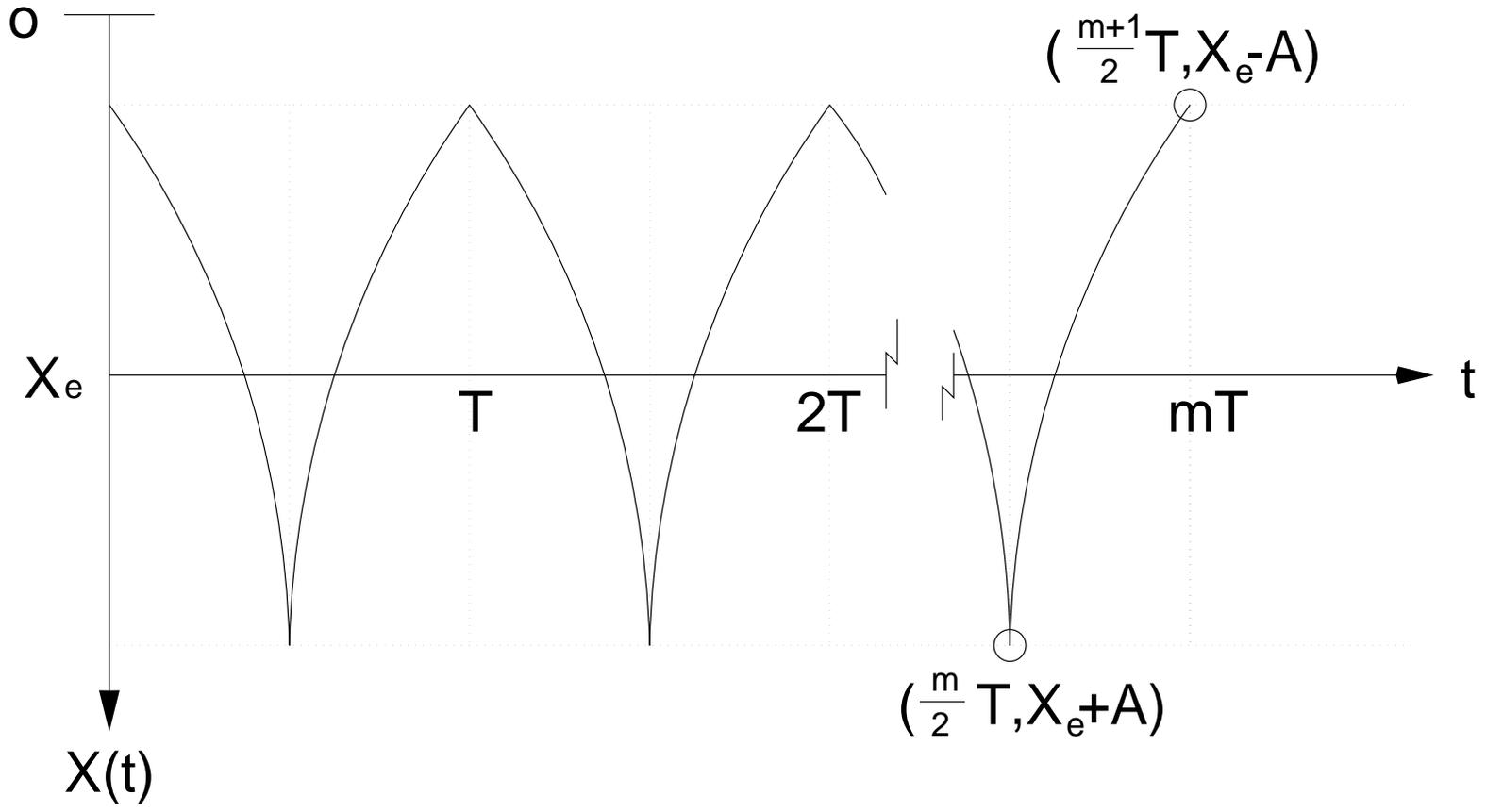
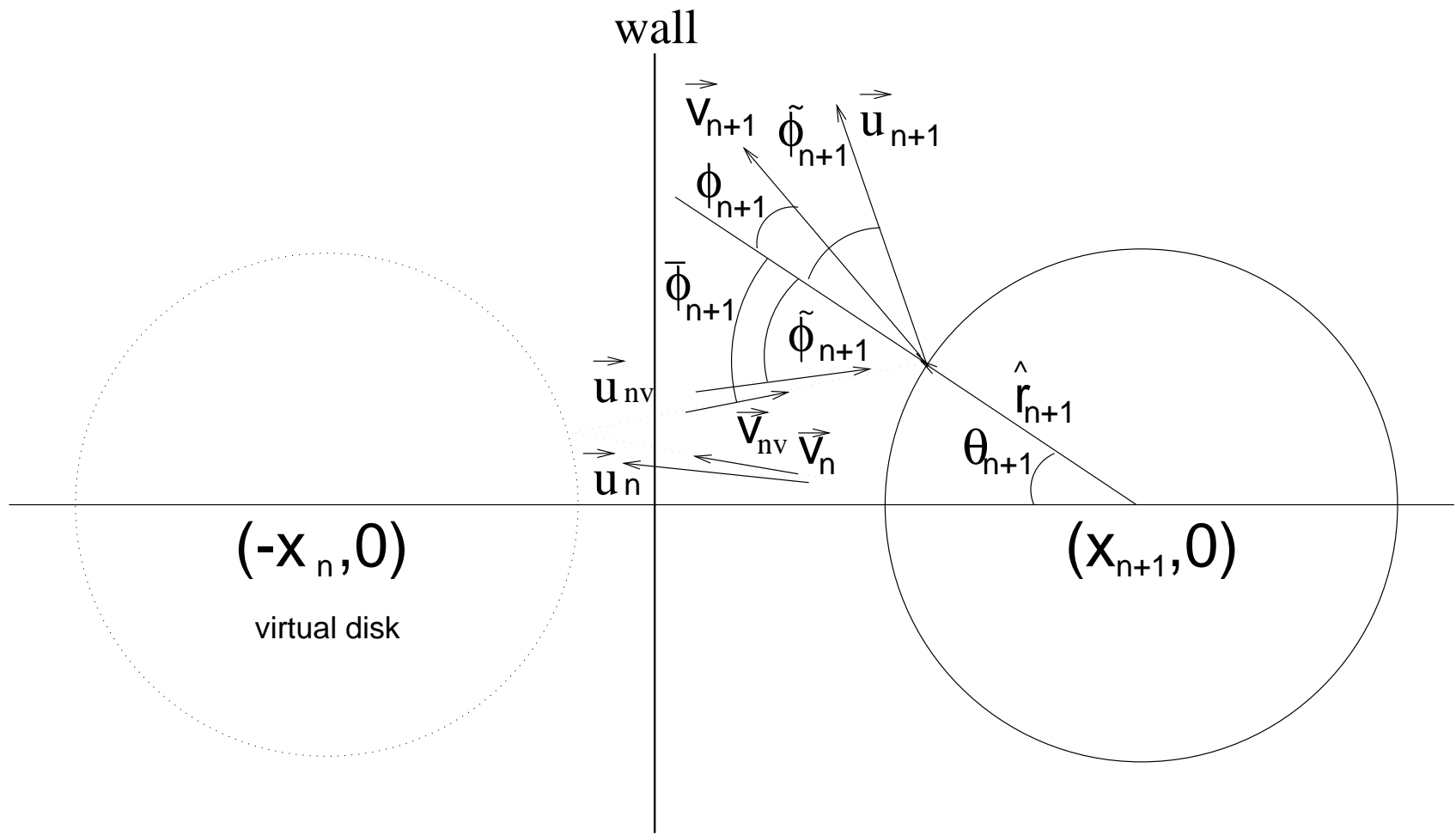
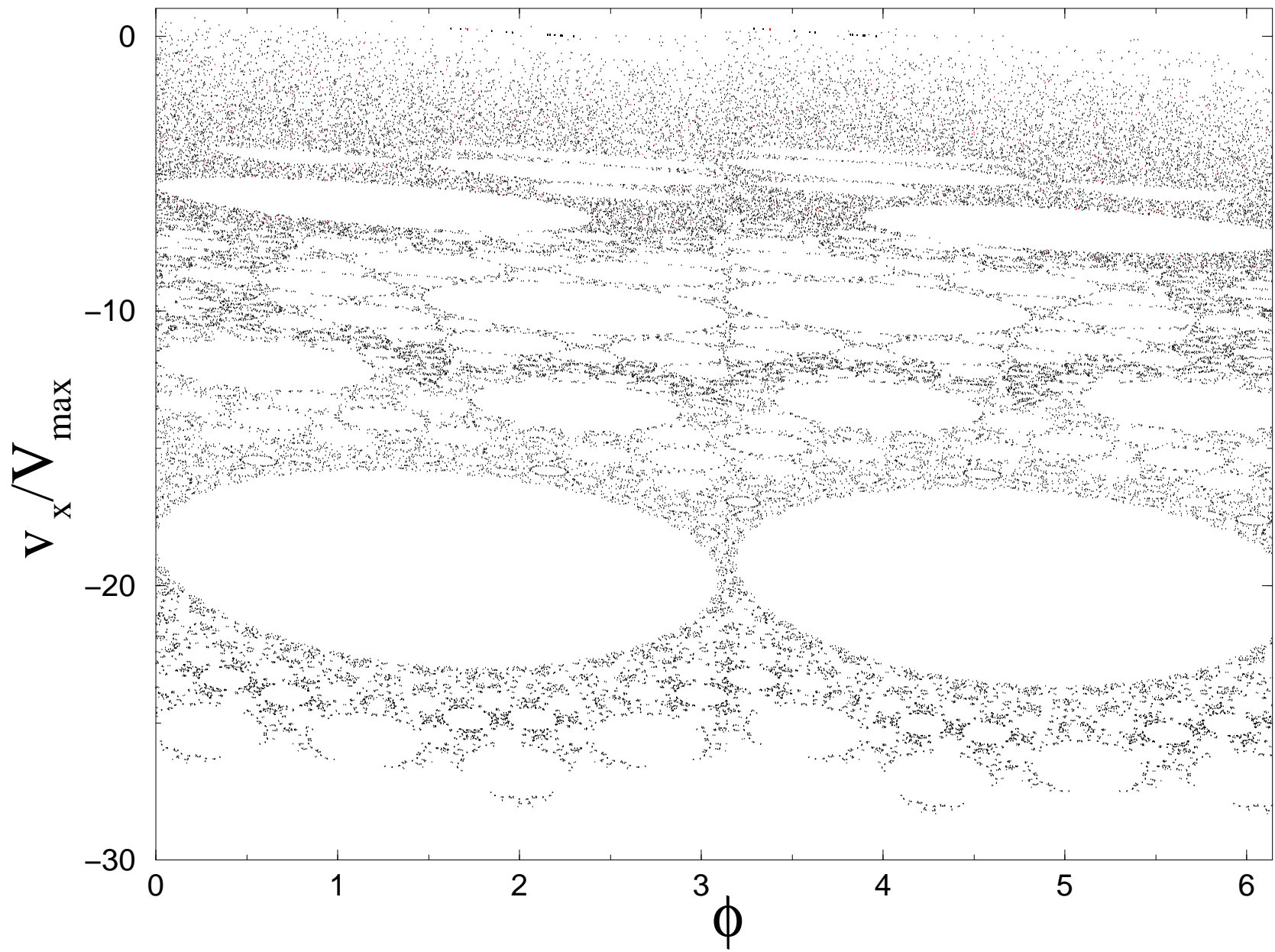
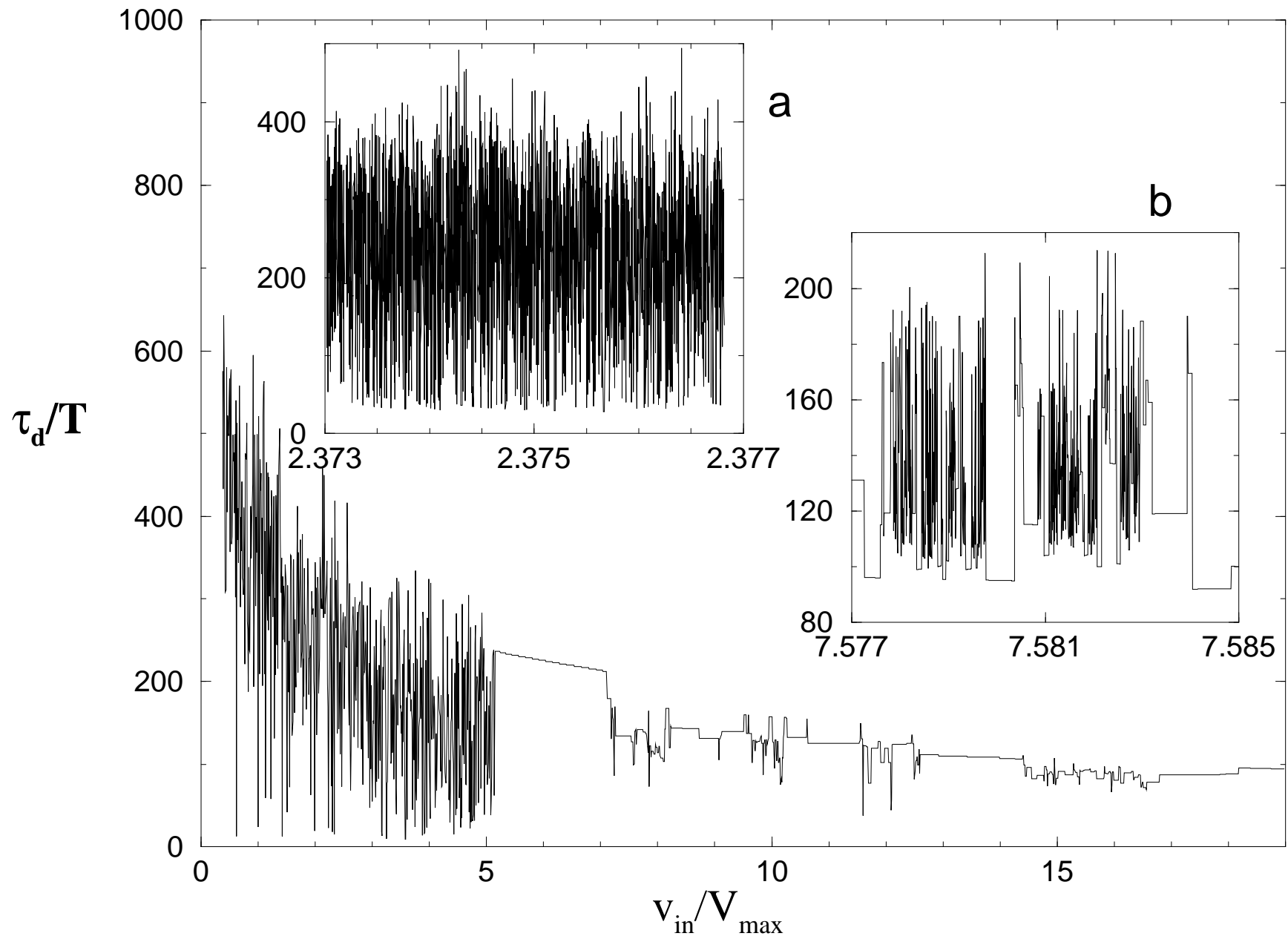
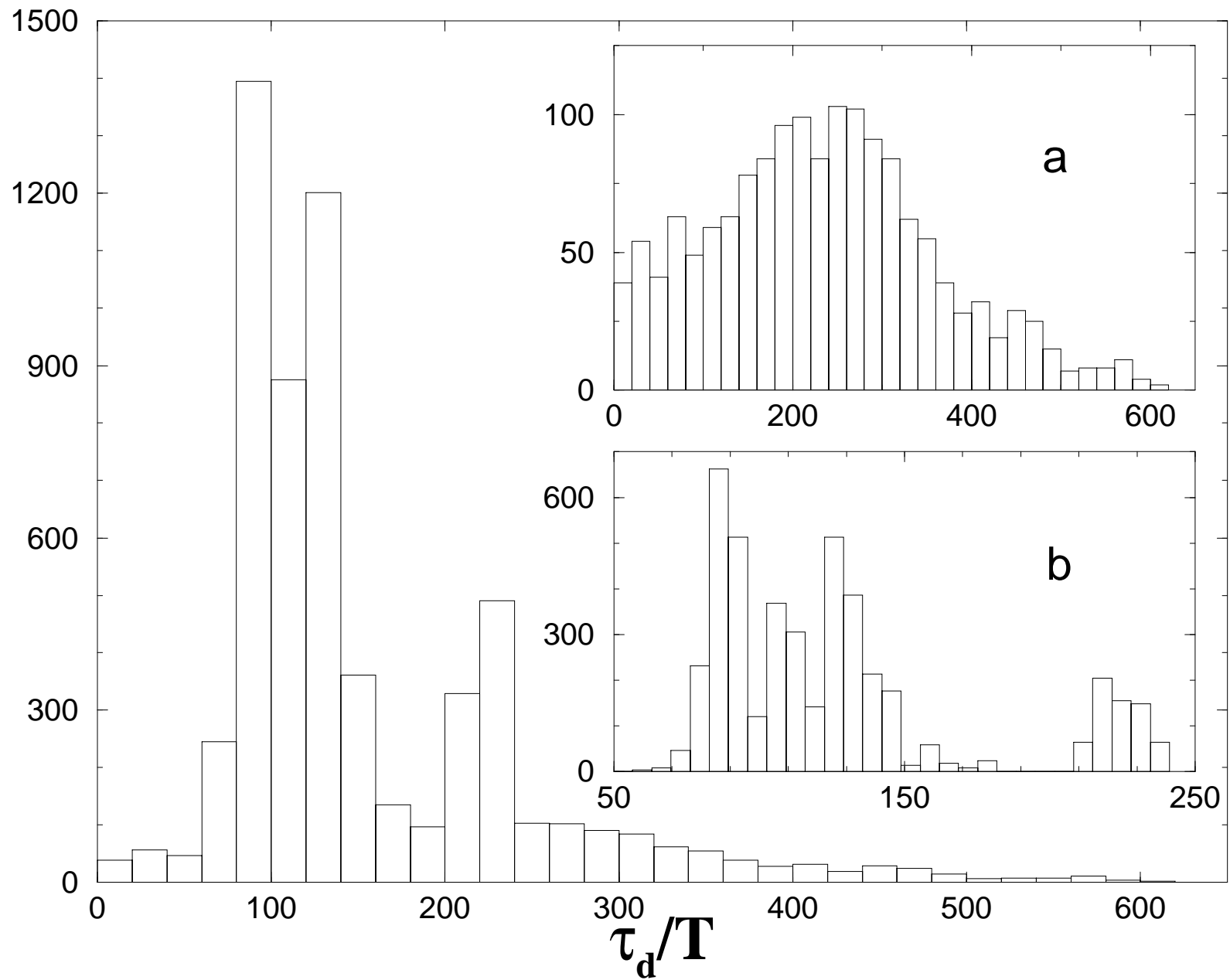
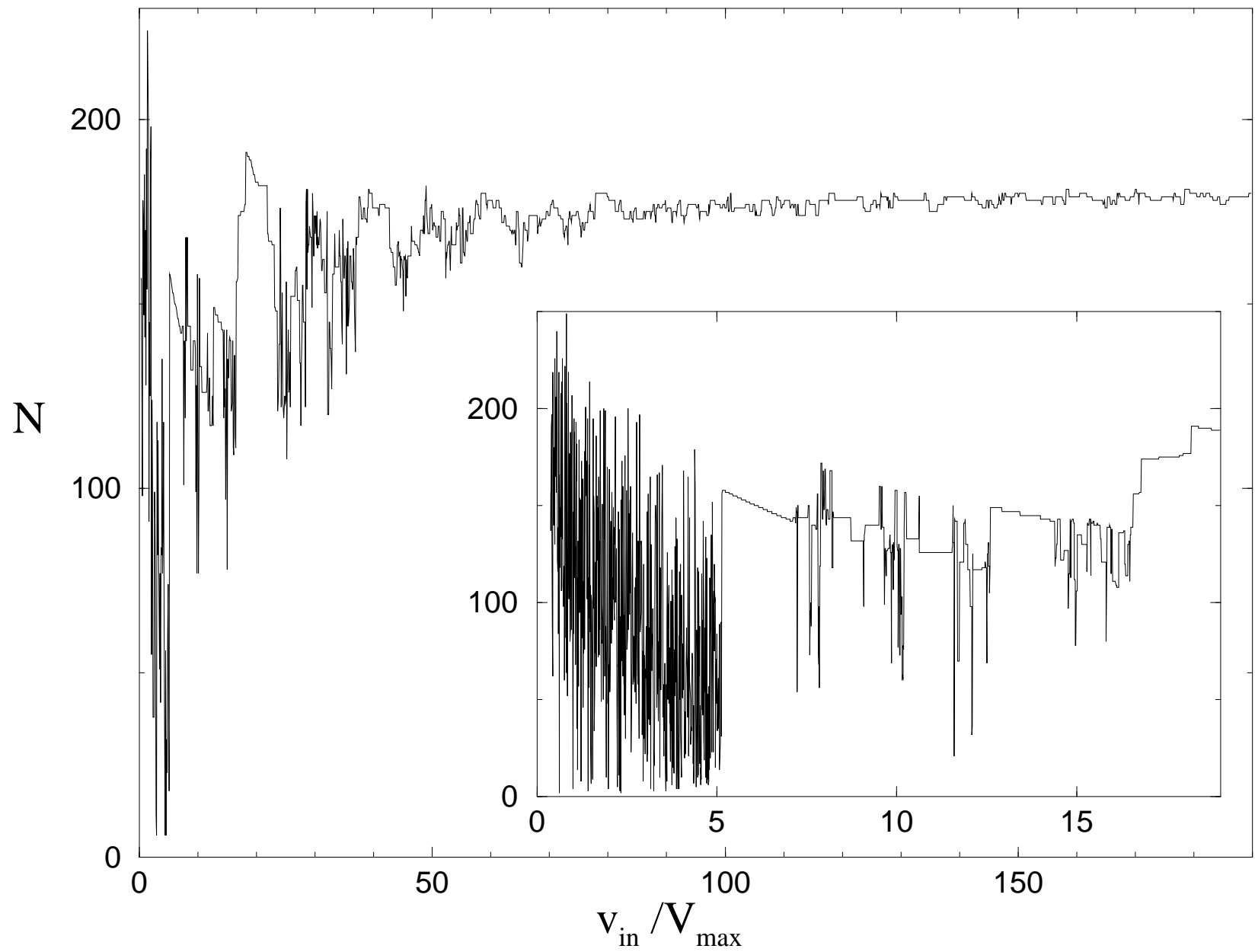
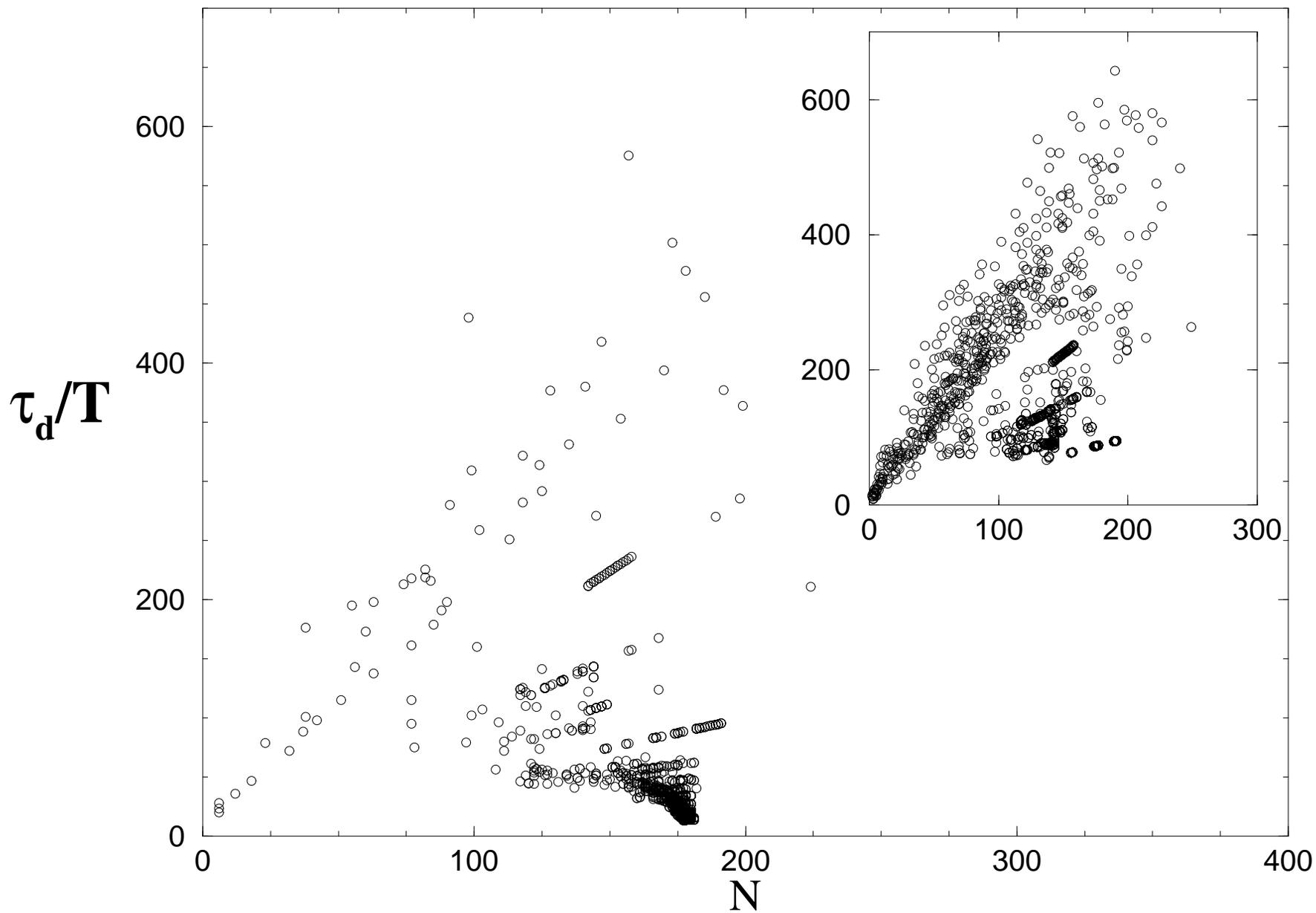
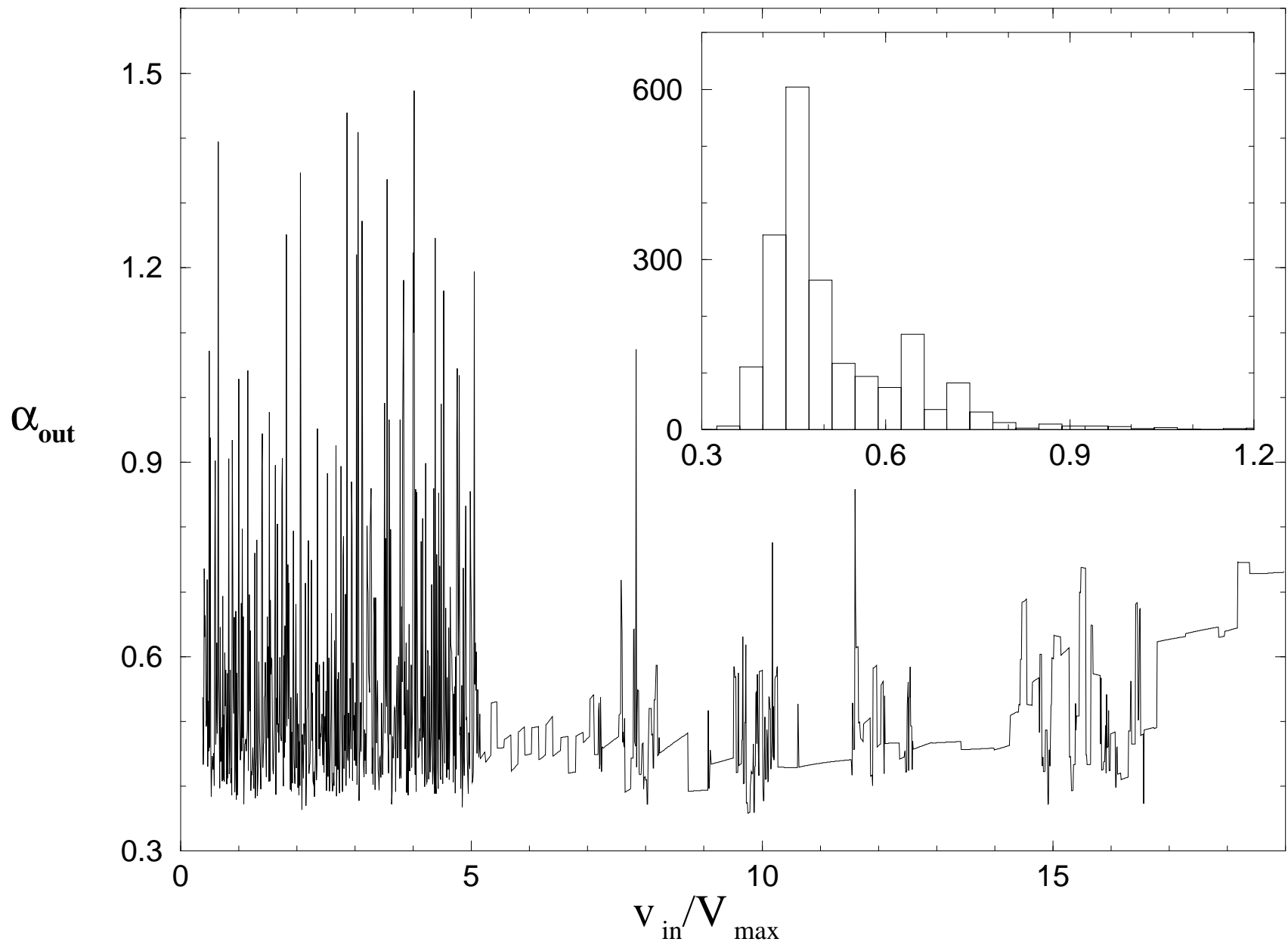
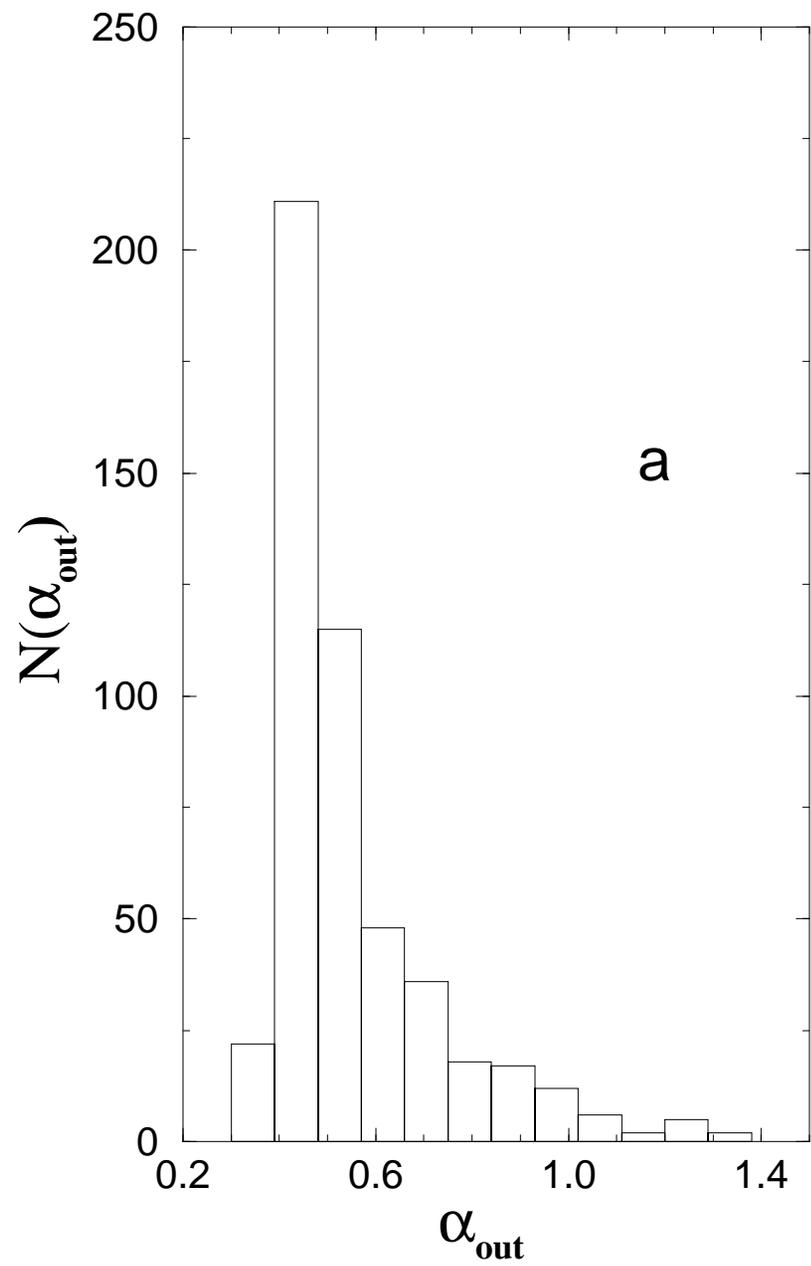
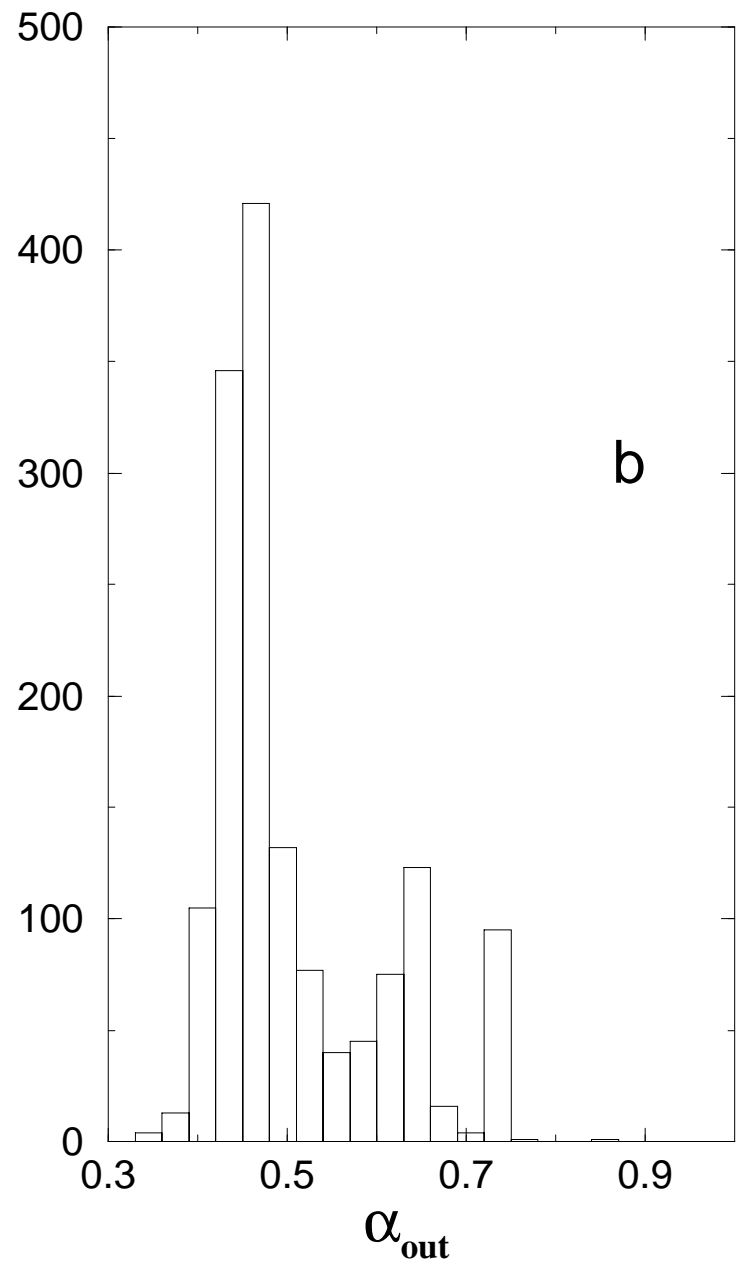
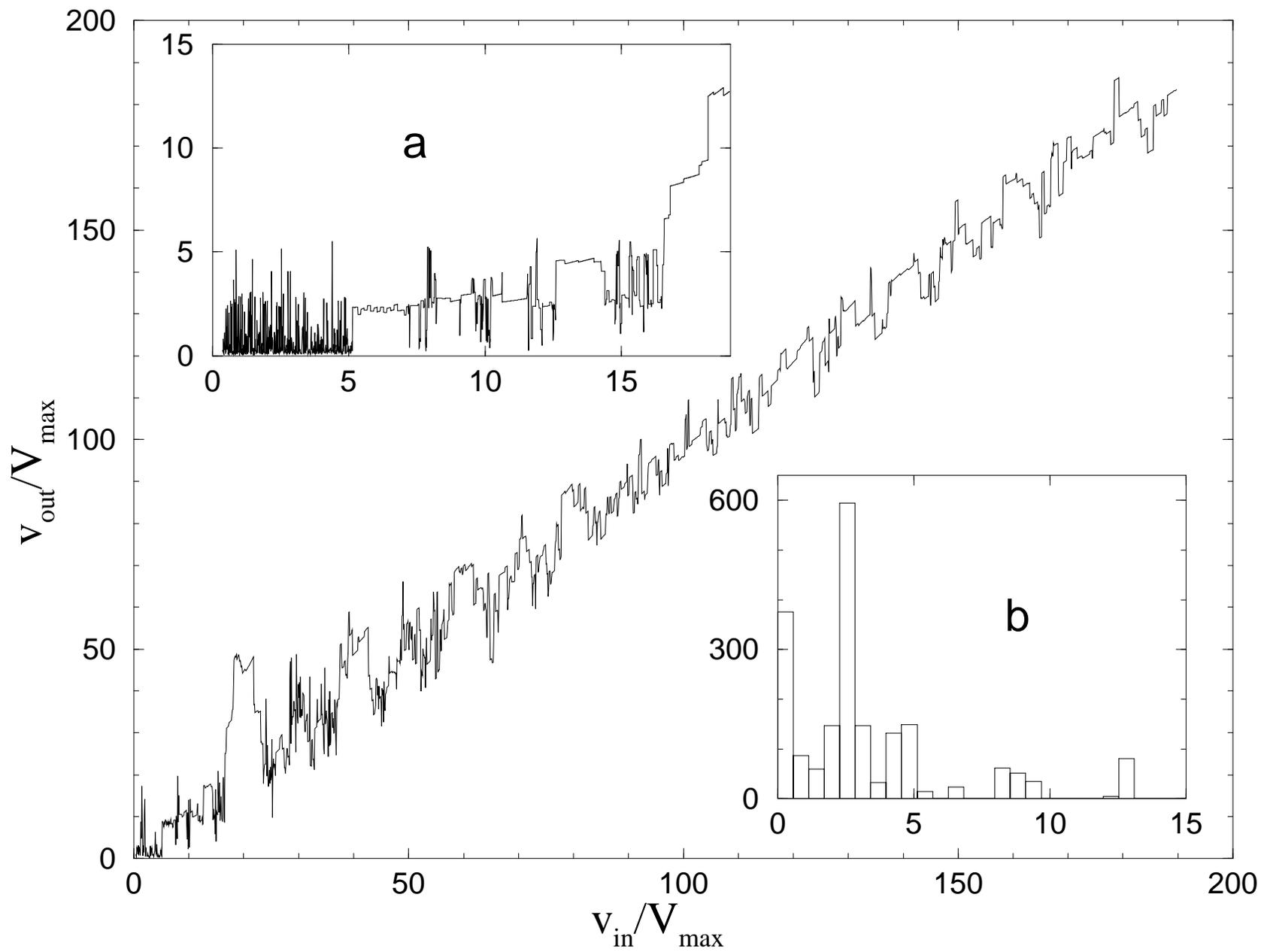
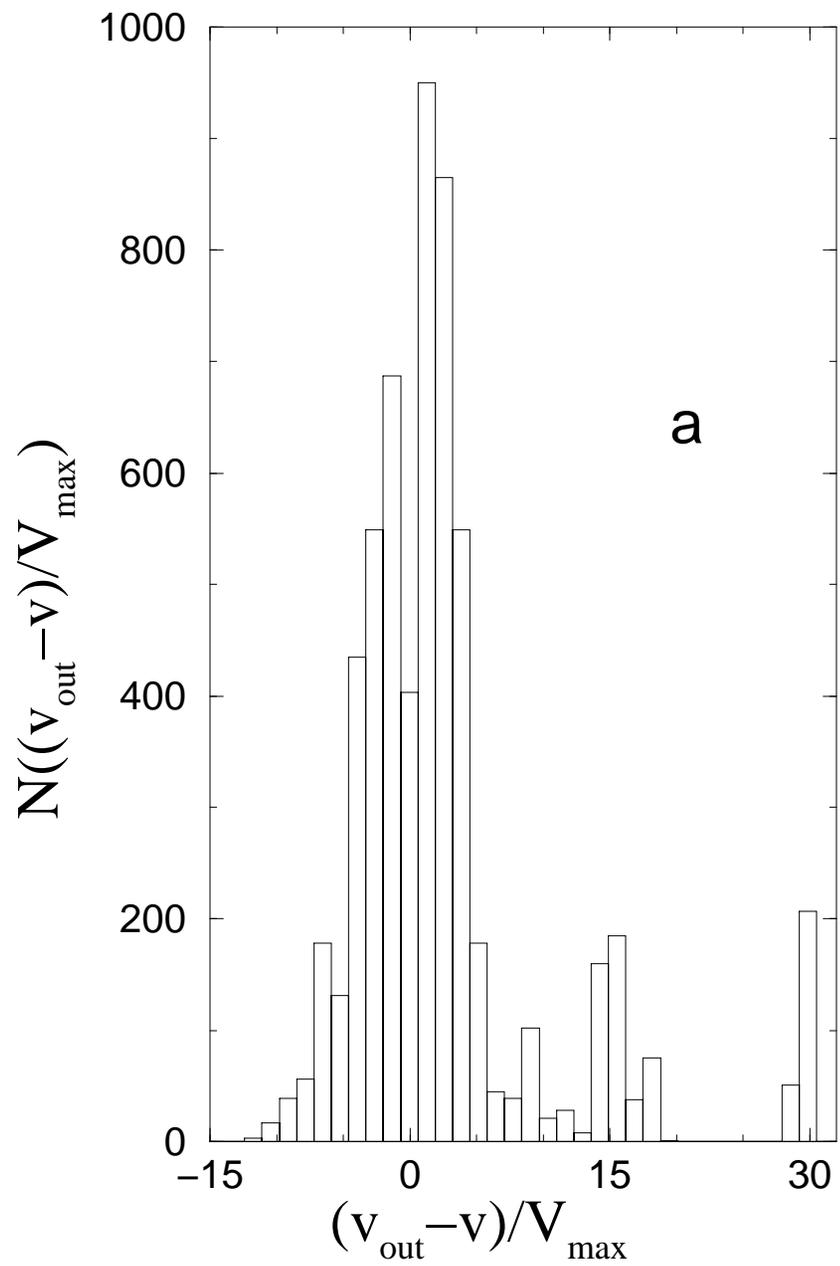
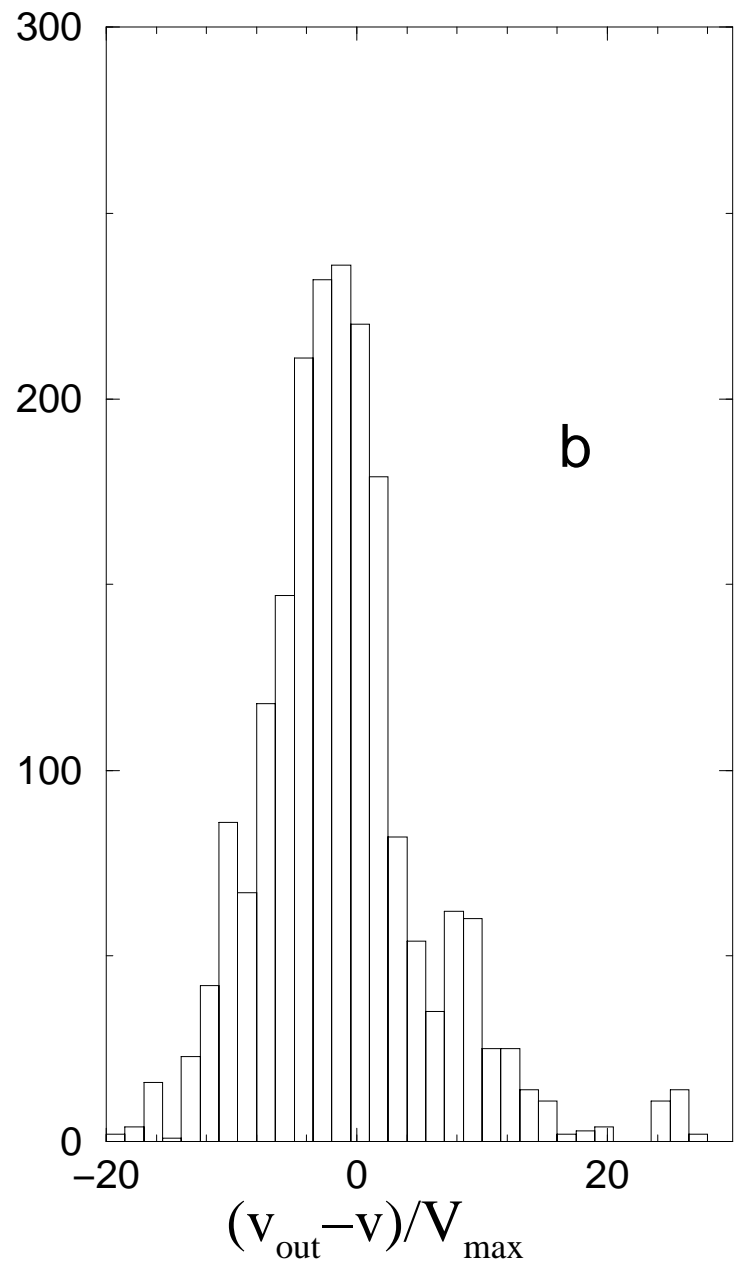
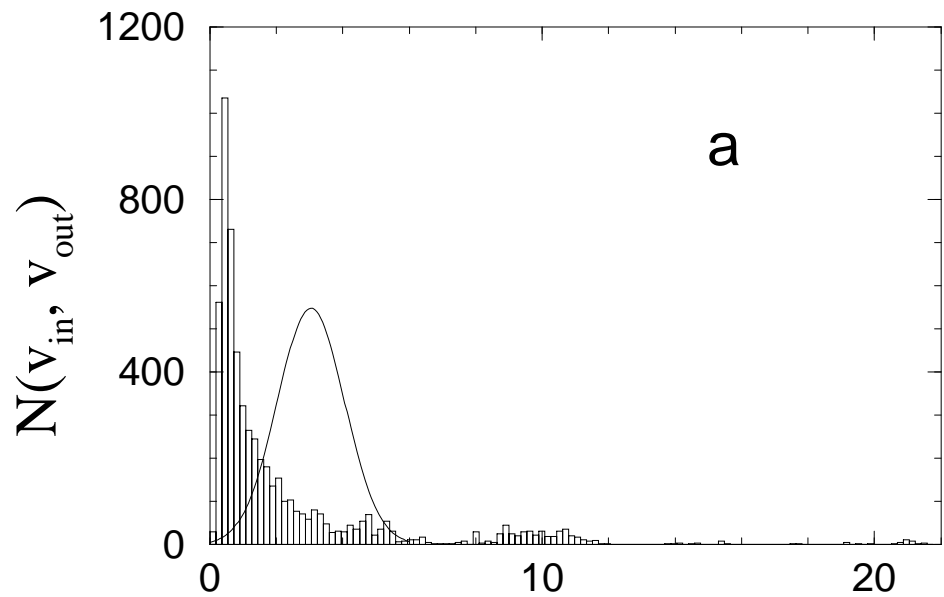
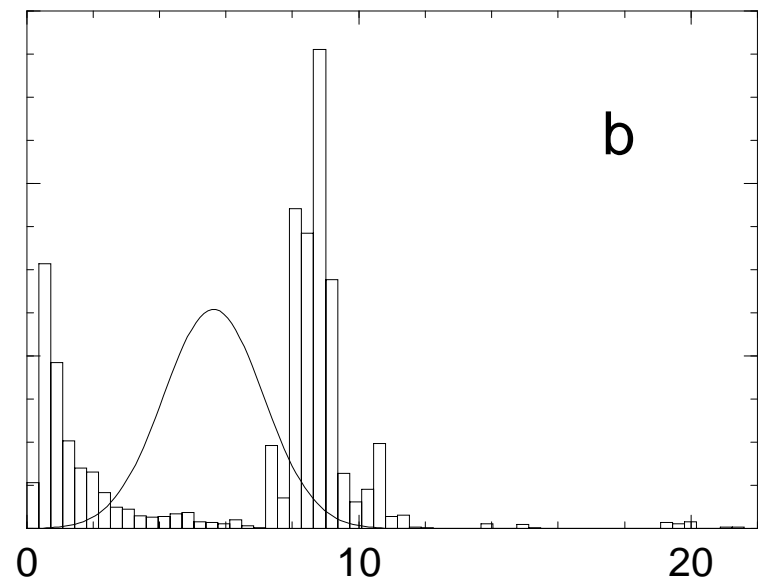
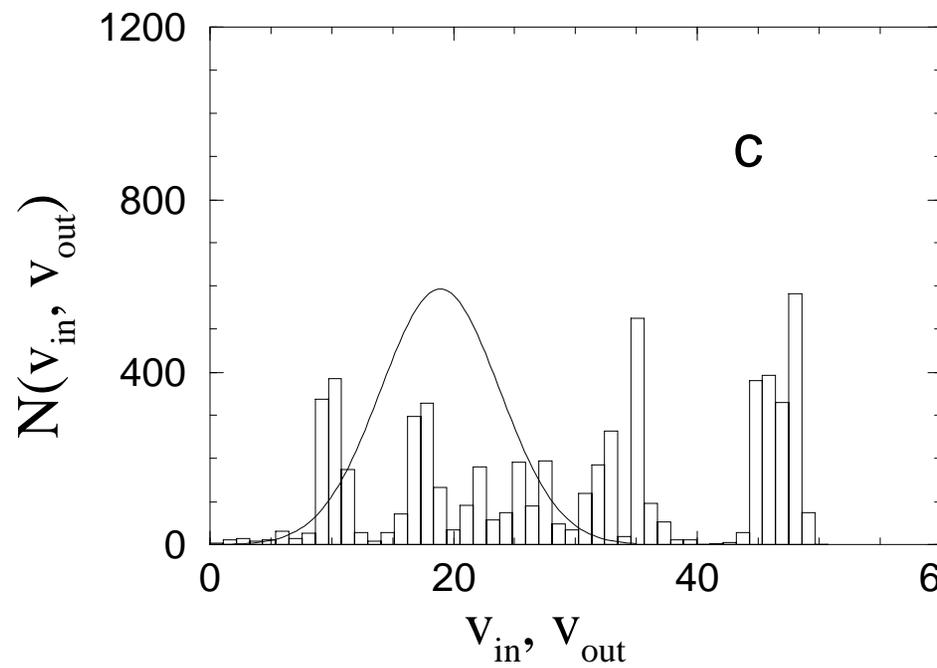
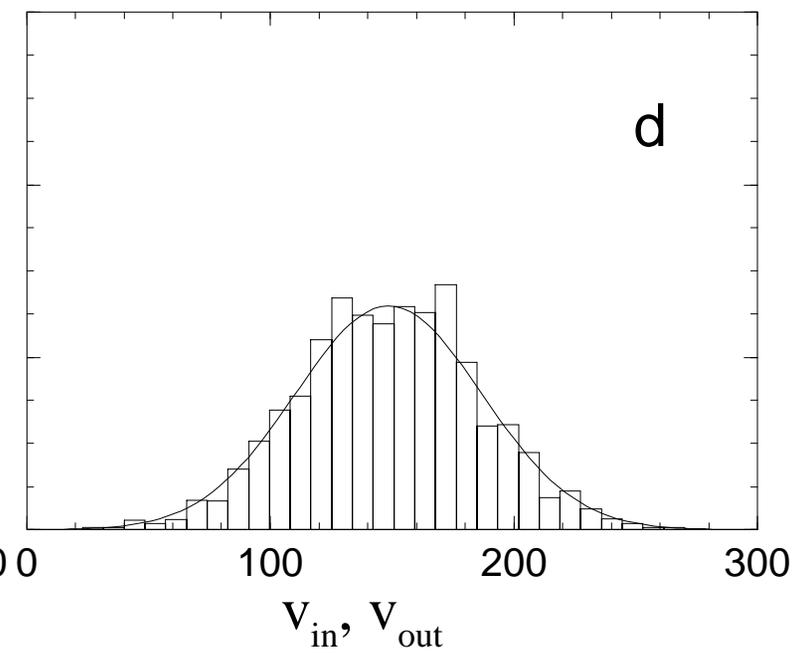
\end{document}